# On the Category of Partial Bijections

**Dr. Emil Schwab, Professor**
„Tibiscus" University of Timisoara, Romania

**ABSTRACT.** Categories of partial functions have become increasingly important principally because of their applications in theoretical computer science. In this note we prove that the category of partial bijections between sets as an inverse-Baer*-category with closed projections and in which the idempotents split is an exact category. Finally the Noether isomorphism theorems are given for this exact category.

## 1 The inverse-Baer*-category of partial bijections

Let X and Y be any two sets. A *partial function* from X to Y is a function from a subset of X called the *domain* of f (which we denote by Dom(f)) to a subset of Y. The *image* of f is the subset Im(f)=f(Dom(f)). For any subset A of X the identity function $1_A$ on A is a partial function from X to itself. We call such partial functions *partial identities*. The empty partial function from X to Y is denoted by $0_{X,Y}$. The composition g ° f of two partial functions f:X→Y and g:Y→Z is given by

$$(g \circ f)(x) = g(f(x)) \quad \text{with} \quad Dom(g \circ f) = \{x \in X \mid f(x) \in Dom(g)\}$$

If Imf∩Domg=ϕ then g ° f = $0_{X,Z}$.

A partial function f which induce a bijection between Def(f) and Im(f) is called a *partial bijection*. If f:X→Y is a partial bijection then the partial bijection $f^{-1}$:Y→X given by

$$Dom(f^{-1}) = \text{Im}(f) \; ; \; \text{Im}(f^{-1}) = Dom(f) \; ; \; f^{-1}(y) = x \text{ iff } f(x) = y$$

173

the *inverse* of f. The basic properties of partial bijections are listed in the following Proposition:

**Proposition 1.1** (i) *The composition of partial bijections is again a partial bijection;*

(ii) *If f:X→Y and g:Y→Z are partial bijections then*

$$f^{-1} \circ f = 1_{Dom(f)} \,;\, f \circ f^{-1} = 1_{Im(f)} \,;\, (f^{-1})^{-1} = f;\, (g \circ f)^{-1} = f^{-1} \circ g^{-1}.$$

(iii) *The set of all partial bijections from a set* X *to itself, denoted* $I(X)$, *is an inverse monoid (called the symmetric inverse monoid on* X*)*

(iv) *The idempotents of* $I(X)$ *are precisely the partial identities on* X.

(v) *If* $1_A, 1_B \in E(I(X))$ *(where* $E(I(X))$ *denote the set of idempotents of* $I(X)$*) then*

$$1_A \circ 1_B = 1_{A \cap B} = 1_B \circ 1_A.$$

(vi) *An inverse semigroup* S *is isomorphic with an inverse subsemigroup of the symmetric inverse monoid* $I(S)$ *of all patial bijections of* S. ■

A category C is called an *inverse category* provided that for each morphism α of C there exists a unique morphism β of C, called the inverse of α, such that

$$\alpha \cdot \beta \cdot \alpha = \alpha \quad and \quad \beta \cdot \alpha \cdot \beta = \beta$$

A great number of results about inverse monoids can be expressed for the inverse categories. We shall verify that a regular category (i.e. a category provided that for each morphism α there exists a morphism β such that α·β·α=α) with the commuting idempotents is an inverse category:

$$\alpha \cdot \beta \cdot \alpha = \alpha \;\Rightarrow\; \alpha \cdot \alpha' \cdot \alpha = \alpha, \alpha' \cdot \alpha \cdot \alpha' = \alpha' \;\; hold \;\; for \;\; \alpha' = \beta \cdot \alpha \cdot \beta$$

If α″ also satisfies

$$\alpha \cdot \alpha'' \cdot \alpha = \alpha \;\; and \;\; \alpha'' \cdot \alpha \cdot \alpha'' = \alpha'',$$

then (note that $\alpha \cdot \alpha', \alpha \cdot \alpha'', \alpha' \cdot \alpha, \alpha'' \cdot \alpha$ are idempotents):

174

$$\alpha' = \alpha' \cdot \alpha \cdot \alpha' = \alpha' \cdot \alpha \cdot \alpha'' \cdot \alpha \cdot \alpha' = \alpha'' \cdot \alpha \cdot \alpha' \cdot \alpha \cdot \alpha' = \alpha'' \cdot \alpha \cdot \alpha'$$

and

$$\alpha'' = \alpha'' \cdot \alpha \cdot \alpha'' = \alpha'' \cdot \alpha \cdot \alpha' \cdot \alpha \cdot \alpha'' = \alpha'' \cdot \alpha \cdot \alpha'' \cdot \alpha \cdot \alpha' = \alpha'' \cdot \alpha \cdot \alpha'$$

and therefore $\alpha' = \alpha''$.

As well as inverse semigroups, the inverse categories are described by subcategories of the category B of partial bijections between sets. Ob(B) is the class of all sets. If X and Y are sets, $\text{Hom}_B(X,Y)$ is the set of all partial bijections from X to Y. The composition of morphisms is the composition of partial bijections. Since for any partial bijection f,

$$f \circ f^{-1} \circ f = f$$

is defined and holds true, the category B is a regular category. Note that

$$\text{Hom}_B(X,X) = I(X)$$

and by Proposition 1.1 (v), the idempotents of B commute. Hence

**Proposition 1.2** *The category* B *of partial bijections between sets is an inverse category.* ∎

As well as the concept of inverse category, the concept of Baer*-category arises if we apply a basic property of Baer*-semigroups (see [ ]) to morphisms of a category with zero. If C is a category, a contravariant endofunctor * on C identical on objects and involutory on morphisms (that is $(\beta \cdot \alpha)^* = \alpha^* \cdot \beta^*$ if the composition $\beta \cdot \alpha$ makes sense, and $(\alpha^*)^* = \alpha$ for any morphism $\alpha$ of C) is called an *involution* on C. An idempotent morphism e of a category C with involution * is called *projection* if e*=e. A category C with involution * and with zero object is called a *Baer*-category* provided that for each morphism $\alpha \in \text{Hom}_C(A,B)$ there exists (a necessarily unique) projection $\alpha' \in \text{Hom}_C(A,A)$ such that

$$\{\beta \in \text{Mor}C \mid \alpha \cdot \beta = 0\} = \alpha' \cdot C,$$

where $\alpha' \cdot C$ denotes the class of all morphisms of the form $\alpha' \cdot \gamma$ ($\gamma \in \text{Mor}C$) such that the composition $\alpha' \cdot \gamma$ makes sense. A projection e of a Baer*-category is called *closed projection* if e″=e.

**Proposition 1.3** *The category* B *of partial bijections between sets is a Baer*-category with closed projections (any projection is closed projecton).*

**Proof.** The category B is provided with a canonical involution, namely $f^*=f^{-1}$. Now, let $f: X \to Y$ be a partial bijection. We shall verify that

$$\{g \in Hom_B(\bullet, X) \mid f \circ g = 0_{\bullet, Y}\} = 1_{X-Def(f)} \circ B \qquad (1_{X-Def(f)} \in Hom_B(X, X))$$

If $g \in Hom(\bullet, X)$ such that $f \circ g = 0_{\bullet, Y}$ then $Im(g) \subseteq X-Def(f)$ and therefore $g = 1_{X-Def(f)} \circ g$. Now, for any morphism $h \in 1_{X-Def(f)} \circ B$ we have $f \circ h = 0_{\bullet, Y}$ because $f \circ 1_{X-Def(f)} = 0_{X,Y}$.

Since the idempotents of B are partial identities, it follows that any idempotent of B is closed projection. ∎

If C is an inverse category then we denote the inverse of $\alpha$ by $\alpha^{-1}$. The inverse category C is provided with the canonical involution $^{-1}$. We say that C is an *inverse -Baer\*-category* if C is an inverse and a Baer\*-category with $* = {}^{-1}$. We have:

**Theorem 1.4** *An inverse- Baer\*-category with closed projections in which every idempotent has a mono-epi factorization is an exact category (in the sense of Mitchell* [MIT65]).

**Proof.** An exact category is a normal and conormal category with kernels and cokernels in which any morphism has a mono-epi factorization. Now, let C be an inverse Baer\*-category with closed projections in which any idempotent has a mono-epi factorization. We shall verify that C is an exact category.

If $\alpha$ is a morphism of C and $\alpha' = p_1 \cdot q_1$ is a mono-epi factorization of $\alpha'$ then $\alpha \cdot \alpha' = 0$ and therefore $\alpha \cdot p_1 = 0$. Let $\alpha \cdot \beta = 0$ for some $\beta \in MorC$. Then $\beta \in \alpha' \cdot C$, that is $\beta = p_1 \cdot (q_1 \cdot \gamma)$ for some morphism $\gamma \in MorC$. It follows $p_1 = ker(f)$. Hence C is a category with kernels. Now, if $(\alpha^{-1})' = p_2 \cdot q_2$ is a mono-epi factorization of $(\alpha^{-1})'$ then $q_2 = coker(\alpha)$. Hence C is a category with kernels and cokernels.

To show that C is normal and conormal let $\alpha$ be a monomorphism in C and $\beta$ an epimorphism in C. It is straightforward to check that $\alpha = ker(\alpha^{-1})'$ and $\beta = coker(\beta')$. Hence C is a normal and conormal category.

To show that any morphism $\alpha$ of C has a mono-epi facotization, let $\alpha \cdot \alpha^{-1} = p \cdot q$ be a mono-epi factorization of the idempotent $\alpha \cdot \alpha^{-1}$. Then,

$$\alpha = \alpha \cdot \alpha^{-1} \cdot \alpha = p \cdot (q \cdot \alpha^{-1} \cdot \alpha) = p \cdot \beta,$$

where $\beta = q \cdot \alpha^{-1} \cdot \alpha$. Note that,

$$\alpha \cdot \alpha^{-1} \cdot p \cdot q = \alpha \cdot \alpha^{-1} \cdot \alpha \cdot \alpha^{-1} = \alpha \cdot \alpha^{-1} = p \cdot q$$

and therefore

$$\alpha \cdot \alpha^{-1} \cdot p = p.$$

It follows

$$p = \alpha \cdot \alpha^{-1} \cdot p = p \cdot \beta \cdot \alpha^{-1} \cdot p,$$

that is:

$$1 = \beta \cdot \alpha^{-1} \cdot p$$

Thus $\beta$ is an epimorhism and $\alpha = p \cdot \beta$ is a mono-epi factorization of $\alpha$. ∎

**Corollary 1.5** *The category* B *of partial bijections between sets is an exact category*

**Proof.** To show that the category B of partial bijections between sets is an exact category it is enough by Theorem 1.4 to show that any idempotent of B has a mono-epi factorization. It is straightforward to check that a morphism $f \in \mathrm{Hom}_B(X,Y)$ is a monomorphism if and only if $\mathrm{Def}(f)=X$, and it is an epimorphism if and only if $\mathrm{Im}(f)=Y$. Thus if $1_A \in \mathrm{Hom}_B(X,X)$ is an idempotent ($\mathrm{Dom}(1_A)=A \subseteq X$) then is a mono-epi factorization of $1_A$ ($1_A = g \circ f$), where the partial bijections $f:X \to A$ and $g:A \to X$ are given by:

$$Def(f) = A;\ f(x) = x \quad and \quad Def(g) = A;\ g(x) = x. \qquad \blacksquare$$

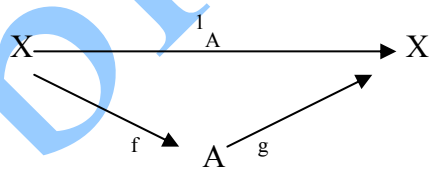

## 2   The Noether isomorphism theorems for the category of partial bijections

Let C be an exact category. The sequence of morphisms in C:

$$0 \longrightarrow U \xrightarrow{\alpha} V \xrightarrow{\beta} W \longrightarrow 0$$

is an exact sequence (called a short exact sequence) if and only if:

$$\alpha = \ker(\beta), \quad (or\ equivalently,\ \beta = \operatorname{coker}(\alpha)) \quad and \quad \begin{cases} \alpha\ is\ a\ monomrphism\ and \\ \beta\ is\ an\ epimorphism \end{cases}$$

In the case of short exact sequence it is frequently denote W by V/U. The undesignated morphism V→V/U in the short exact sequence

$$0 \longrightarrow U \xrightarrow{\alpha} V \longrightarrow V/U \longrightarrow 0,$$

will be understood to be β=coker(α).

The following statement is true in an exact category C:

**Proposition 2.1** (The 3×3 Lemma) *A commutative diagram in an exact category*

$$\begin{array}{ccccccccc}
 & & 0 & & 0 & & 0 & & \\
 & & \downarrow & & \downarrow & & \downarrow & & \\
0 & \to & U' & \to & U & \to & U'' & \to & 0 \\
 & & \downarrow & & \downarrow & & \downarrow & & \\
0 & \to & V' & \to & V & \to & V'' & \to & 0 \\
 & & \downarrow & & \downarrow & & \downarrow & & \\
 & & W' & & W & & W'' & & \\
 & & \downarrow & & \downarrow & & \downarrow & & \\
 & & 0 & & 0 & & 0 & &
\end{array}$$

where all the rows and columns are exact, can be completed by an exact sequence

$$0 \longrightarrow W' \longrightarrow W \longrightarrow W'' \longrightarrow 0.$$

In what follows the notation U⊆V signifies a fixed monomorphism U→V, hence an exact sequence

$$0 \longrightarrow U \longrightarrow V \longrightarrow V/U \longrightarrow 0.$$

Let U⊆V⊆W. Hence the diagram

$$\begin{array}{ccccccc}
& & 0 & & 0 & & 0 \\
& & \downarrow & & \downarrow & & \downarrow \\
0 & \to & U & \Rightarrow & U & \to & 0 & \to & 0 \\
& & \downarrow & & \downarrow & & \downarrow \\
0 & \to & V & \to & W & \to & W/V & \to & 0 \\
& & \downarrow & & \downarrow & & \Downarrow \\
& & V/U & & W/U & & W/V \\
& & \downarrow & & \downarrow & & \downarrow \\
& & 0 & & 0 & & 0
\end{array}$$

(where $\Rightarrow$ is the identity morphism) is a commutative diagram where all the rows and columns are exact. By Proposition 2.1 it follows:

**Theorem 2.2** (First Noether isomorphism theorem for exact category) *Let* $U \subseteq V \subseteq W$ *in an exact category. Then*

$$V/U \subseteq W/U \quad and \quad (W/U)/(V/U) \cong W/V. \qquad \blacksquare$$

The second Noether isomorphism theorem for exact category is given by

**Theorem 2.3** (Second Noether isomorphism theorem for exact category) *Let* $U \subseteq W$ *and* $V \subseteq W$ *in an exact category. Then*

$$V/U \cap V \cong U \cup V/V. \qquad \blacksquare$$

The purpose of this section is to find the Noether set-theoretical identities that is an illustration of the last two theorems in a non-abelian category namely in the exact (and balanced) category B of partial bijections between sets. Since a morphism $f \in Hom_B(X,Y)$ is a monomorphism (epimorphism) if and only if $Def(f)=X$ ($Imf=Y$), it follows that if f is an isomorphism than card(X)=card(Y). So, the subobjects of an object $X \in Ob(B)$ are the subsets of the set X. The notation $X_1 \subseteq X$ is suggestive for a subobject of X. If the sequence

$$0 \longrightarrow X_1 \longrightarrow X \longrightarrow Y \longrightarrow 0$$

is exact in B, then card(Y)=card(X–$X_1$). Thus the quotient object Y of X will be understood to be X-$X_1$. The short exact sequences in B are the sequences

$$0 \longrightarrow X_1 \longrightarrow X \longrightarrow X - X_1 \longrightarrow 0$$

with $X_1$ subset of X.

The Noether isomorphism theorems for the category B of partial bijections between sets becomes:

**Theorem 2.4** ( First Noether isomorphism theorem for the category of partial bijections between sets) *Let* $X_1 \subseteq X_2 \subseteq X$. *Then*

$$X_2 - X_1 \subseteq X - X_1 \quad and \quad (X - X_1) - (X_2 - X_1) \equiv X - X_2 \qquad \blacksquare$$

**Theorem 2.5** (Second Noether isomorphism theorem for the category of partial bijections between sets) *Let* $X_1, X_2 \subseteq X$. *Then*

$$X_2 - (X_1 \cap X_2) \equiv (X_1 \cup X_2) - X_1. \qquad \blacksquare$$